\documentclass[letterpaper,twocolumn,pra,
aps,showpacs,superscriptaddress,floatfix]{revtex4-2}

\usepackage[utf8]{inputenc}
\usepackage{bbm}
\usepackage{bm}
\usepackage[usenames]{color}
\usepackage{multirow}
\usepackage{amssymb}
\usepackage{amsbsy}
\usepackage{mathtools}
\usepackage{amsmath}
\usepackage{stmaryrd}
\usepackage{graphicx}
\usepackage{epsfig}
\usepackage{placeins}
\usepackage{bbold}
\usepackage{braket}
\usepackage{blindtext}
\usepackage[colorlinks,linkcolor=blue,citecolor=blue,urlcolor=blue]{hyperref}

\usepackage{scalerel}
\usepackage{tikz}
\usetikzlibrary{calc}
\usetikzlibrary{patterns}
\usetikzlibrary{svg.path}
\definecolor{orcidlogocol}{HTML}{A6CE39}
\tikzset{
  orcidlogo/.pic={
    \fill[orcidlogocol] svg{M256,128c0,70.7-57.3,128-128,128C57.3,256,0,198.7,0,128C0,57.3,57.3,0,128,0C198.7,0,256,57.3,256,128z};
    \fill[white] svg{M86.3,186.2H70.9V79.1h15.4v48.4V186.2z}
                 svg{M108.9,79.1h41.6c39.6,0,57,28.3,57,53.6c0,27.5-21.5,53.6-56.8,53.6h-41.8V79.1z M124.3,172.4h24.5c34.9,0,42.9-26.5,42.9-39.7c0-21.5-13.7-39.7-43.7-39.7h-23.7V172.4z}
                 svg{M88.7,56.8c0,5.5-4.5,10.1-10.1,10.1c-5.6,0-10.1-4.6-10.1-10.1c0-5.6,4.5-10.1,10.1-10.1C84.2,46.7,88.7,51.3,88.7,56.8z};
  }
}
\newcommand{\new}[1]{\textcolor{black}{#1}}
\newcommand{\jz}[1]{\textcolor{black}{#1}}

\newcommand\orcid[1]{\!%
  \href{https://orcid.org/#1}{%
    \mbox{%
      \scaleto{%
        \begin{tikzpicture}[yscale=-1,transform shape]
          \pic{orcidlogo};
        \end{tikzpicture}
      }{8pt}%
    }%
  }%
}
\makeatletter

\begin{document}
\title{Random matrix theory universality of  current operators in  spin-$S$ Heisenberg chains}

\author{Mariel Kempa~\orcid{0009-0006-0862-4223}}
\email{makempa@uos.de}
\affiliation{University of Osnabr{\"u}ck, Department of Mathematics/Computer
	Science/Physics, D-49076 Osnabr{\"u}ck, Germany}

\author{Markus Kraft~\orcid{0009-0008-4711-5549}}
\affiliation{University of Osnabr{\"u}ck, Department of Mathematics/Computer
	Science/Physics, D-49076 Osnabr{\"u}ck, Germany}

\author{Robin Steinigeweg~\orcid{0000-0003-0608-0884}}
\email{rsteinig@uos.de}
\affiliation{University of Osnabr{\"u}ck, Department of Mathematics/Computer
	Science/Physics, D-49076 Osnabr{\"u}ck, Germany}

\author{Jochen Gemmer~\orcid{0000-0002-4264-8548}}
\affiliation{University of Osnabr{\"u}ck, Department of Mathematics/Computer
	Science/Physics, D-49076 Osnabr{\"u}ck, Germany}

\author{Jiaozi Wang~\orcid{0000-0001-6308-1950}}
\email{jiaowang@uos.de}
\affiliation{University of Osnabr{\"u}ck, Department of Mathematics/Computer
	Science/Physics, D-49076 Osnabr{\"u}ck, Germany}

\date{\today}
%------------------------------------------------------------------------------
% Abstract
%------------------------------------------------------------------------------
\begin{abstract}
Quantum chaotic systems exhibit certain universal statistical properties that closely resemble predictions from random matrix theory (RMT).
With respect to observables, it has recently been conjectured that, when truncated to a sufficiently narrow energy window, their statistical properties can be described by an unitarily invariant ensemble, and testable criteria have been introduced, which are based on the scaling behavior of free cumulants.
In this paper, we investigate the conjecture numerically in translationally invariant Heisenberg spin chains with spin quantum number $S =\frac{1}{2},1,\frac{3}{2}$. 
Combining a quantum-typicality-based numerical method with the exploitation of the system's symmetries,  we study the spin current operator and find clear evidence of consistency with the proposed criteria in chaotic cases. Our findings further support the conjecture of the existence of RMT universality as manifest in the observable properties in quantum chaotic systems.

\end{abstract}

\maketitle

%------------------------------------------------------------------------------
% Introduction
%------------------------------------------------------------------------------
\section{Introduction}

Understanding how quantum many-body systems thermalize is one of the most fundamental and challenging problems in quantum physics.
During the last decades, with the introduction and development of the concept of quantum chaos \cite{Haake-chaos-book,GUHR1998189-chaos-review,IZRAILEV1990299-chaos-review,casati2006quantum-book-chaos,ZELEVINSKY199685-chaos-review,RevModPhys.53.385-chaos-review}, the understanding of this question has been substantially improved.

\new{The most general defining feature of quantum chaos is the emergence of universal statistical properties consistent with the predictions of random matrix theory (RMT) \cite{GUHR1998189-chaos-review,Haake-chaos-book,dkpr-16}(see also Ref.~\cite{Richter_2022-chaos-rmt} for a recent review).}
Among the most well-known signatures are the spectral properties. It is conjectured that the spectral correlations of a quantum system with a chaotic classical limit follow the description of RMT \cite{Bohigas,Casati-chaos80}. This conjecture has been further supported and established on a firmer theoretical footing through semiclassical analysis \cite{Berry-chaos-81,Casati-chaos80,Berry-chaos-85,Sieber-chaos-01,Haake-chaos-book,Kaufman-chaos-79,Mueller-chaos-04,Mueller-chaos-05,Wigner-chaos} and, more recently, also in several quantum many-body systems without a clear defined classical limit \cite{PhysRevX.8.021062_RMT_Prosen,PhysRevLett.121.264101_RMT_Prosen, PhysRevX.8.041019-chaos-sff}.

Besides spectral properties, statistical properties of energy eigenstates in quantum chaotic systems also show universal behavior. In the configuration basis, Berry's conjecture \cite{berry1977} postulates components of chaotic eigenstates as Gaussian random variables, consistent with the prediction of RMT. Extensions of this conjecture to more general bases, such as the Fock basis, or the integrable basis, have also been studied \cite{PhysRevE.97.062219-wang-ef-chaos, PhysRevLett.134.010404-RMT-EF}.

\new{In addition to spectra and eigenstates, observables are also expected to exhibit certain universal statistical properties. 
The eigenstate thermalization hypothesis (ETH) \cite{Srednicki94,Deutsch-91,Srednicki_1996,Srednicki1999,rigol2008thermalization,dkpr-16} provides a natural framework for understanding these properties by postulating the structure of observable matrix elements in the energy eigenbasis. 
The ETH ansatz suggests that, when truncated to a sufficiently small energy window, the statistical properties of an operator can be described by a random matrice drawn from the Gaussian unitary ensemble(GUE) (or Gaussian orthogonal ensemble(GOE)).
However, more recent studies have revealed that correlations among matrix elements persist for any nonzero energy window\cite{Dymarsky22-ETH-offdiag, Wang22-ETH-offdiag, Jonas-offdiag, iniguez2023microcanonical,PhysRevLett.123.260601-uie-offdiag, wang2024emergence-uie-offdiag, vallini2025refinements-offdiag-feth, PhysRevLett.123.260601-uie-offdiag}, thereby invalidating the simple GUE picture described above.}
  Motivated by the full eigenstate thermalization hypothesis \cite{GETH-Foini19,PhysRevLett.134.140404-feth-fp,PhysRevLett.129.170603-feth-fp}, it has been conjectured that \cite{wang2024emergence-uie-offdiag,PhysRevLett.123.260601-uie-offdiag}, within sufficiently narrow microcanonical energy windows, truncated operators 
can be effectively described by a sample of unitarily invariant ensemble. 
Numerical evidence has been found in a mixed-field spin $S = \frac{1}{2}$ chain \cite{wang2024emergence-uie-offdiag}, as well as in a few-body (two-body) system \cite{wang2025eigenstate-uie-classical} with a chaotic classical limit. However, the existence of the conjectured RMT universality in more generic systems remains an open question. In particular, spin chains with higher spin quantum number $S$ have not been explored, mainly due to the rapidly increasing numerical complexity with increasing 
$S$.

In this work, we take a first step and extend the investigation to a translationally invariant Heisenberg chains with spin quantum number $S = \frac{1}{2},1,\frac{3}{2}$. To mitigate the increasing numerical cost, we combine a previously developed quantum-typicality-based numerical approach \cite{Wang22-ETH-offdiag,Robin_DQPT_2014,fuellgraf2025scalingfreecumulantsclosed-typ, Robin_diffusion_2015, Robin_diffusion_2017,PhysRevLett.102.110403-DQT,Jin2021,HeitmannRichterSchubertSteinigeweg-DQT} with the exploitation of the system's symmetries, which enables us to reach Hilbert space dimensions far beyond the scope of exact diagonalization.
We find numerical evidence that,
when being truncated to a sufficiently narrow energy window, 
statistical properties of the spin current operator follow the description of a unitarily invariant ensemble, provided that the system is chaotic.
These results support the conjecture that RMT universality in observable statistics is a generic feature of chaotic quantum many-body systems.

The remainder of this paper is organized as follows. Sec.~\ref{frame} introduces the theoretical framework. Sec.~\ref{numerical-section} presents the numerical investigation and is organized into three subsections: the models and observables (\ref{model}), the numerical methods (\ref{method}), and the numerical results (\ref{numerics}). Conclusion and discussion are given in Sec.~\ref{conclusion}.

%------------------------------------------------------------------------------
% Framework
%------------------------------------------------------------------------------
\section{Framework}\label{frame}
In this section, we present the theoretical framework underlying our study, namely the RMT universality of microcanonically truncated operators.  While parts of the content in this section have been previously presented in Ref.~\cite{wang2024emergence-uie-offdiag}, here we instead focus on the conceptual structure, including  motivation, reasoning, and  physical insights, with the aim of providing a more concise and intuitive understanding.

\subsection{Microcanonically truncated operator as a sample of unitarily invariant ensemble}
Let us consider a quantum many-body system, the Hamiltonian of which is denoted as $H$. 
The eigenstates and eigenvalues satisfy
\begin{equation}
    H | m \rangle = E_m | m \rangle ,
\end{equation}
where the eigenvalues $E_m$ are labeled in ascending order.
We now take a few-body observable of interest, $O$, and project it onto an energy window,
\begin{equation}
\label{OdE}
    O_{\Delta E} = P_{\Delta E}\, O\, P_{\Delta E} .
\end{equation}
Here $P_{\Delta E}$ denotes the microcanonical projection operator,
\begin{equation}
\label{Eq::EFilter}
    P_{\Delta E} = \sum_{|E_m - E_0| < \Delta E/2} \ket{m}\bra{m} ,
\end{equation}
where $\Delta E$ and $E_0$ are the width and center of the energy window, and
we use $d_{\Delta E}={\rm Tr}(P_{\Delta E})$ to denote the number of states within the microcanonical window. 

\new{Before entering into a more detailed discussion, it is useful to clarify the physical relevance of the truncated observable defined above. 
To this end, let us consider a generic out-of-equilibrium initial state 
\begin{equation}
|\psi(0)\rangle = \sum_m c_m |m\rangle ,
\end{equation}
where $|m\rangle$ denotes an eigenstate of the Hamiltonian with eigenvalue $E_m$. 
The time evolution of the expectation value of an observable $O$ is then given by
\begin{equation}
\langle O(t)\rangle =
\sum_{mn} c_m^{*} c_n O_{mn} e^{-i(E_n-E_m)t} .
\end{equation}
At $t=0$, the phase factor is unity, $e^{-i(E_n-E_m)t}=1$, and therefore all matrix elements $O_{mn}$ contribute to $\langle O(t)\rangle$. 
By contrast, at a finite time $t=T$, matrix elements with energy differences much larger than the inverse time scale, 
$|E_n-E_m| \gg 2\pi/T$, acquire rapidly oscillating phases and tend to dephase. 
As a result, the dynamics on the time scale $T$ is mainly dependent on matrix elements satisfying
\begin{equation}
|E_n-E_m| \lesssim \frac{2\pi}{T}.
\end{equation}
This provides a physical motivation for introducing the truncated observable, 
in the sense that the statistical properties of ${O}_{\Delta E}$ encode information about the out-of-equilibrium dynamics of $O$ on time scales 
$T \sim 2\pi/\Delta E$ \cite{Dymarsky22-ETH-offdiag}.}

\new{For large energy windows $\Delta E$, ${O}_{\Delta E}$ probes short-time dynamics, whose behavior is typically non-universal and strongly dependent on the choice of observable. 
Consequently, universal properties of ${O}_{\Delta E}$ in this regime are general not expected.
For small energy windows $\Delta E$, on the other hand, ${O}_{\Delta E}$ probes dynamics on longer time scales. 
In chaotic systems, few-body observables are expected to thermalize after a characteristic time scale, referred to as the thermalization time $T_{\text{eq}}$. 
Therefore, once the corresponding time scale $2\pi/\Delta E$ exceeds $T_{\text{eq}}$, or equivalently when 
$\Delta E \lesssim \Delta E_{\text{eq}} \sim \frac{2\pi}{T_{\text{eq}}}$,
one may expect the statistical properties of ${O}_{\Delta E}$ to exhibit universal behavior.
}

The picture proposed above is consistent with the ETH ansatz \cite{Srednicki94,Deutsch-91,Srednicki_1996,Srednicki1999,rigol2008thermalization,dkpr-16}, which conjectures the following form of ${O}_{mn} = \langle m|{O}|n\rangle$:
\begin{equation}\label{eq::ETH}
 {O}_{mn} = O(\bar{E})\delta_{mn} + 
\Omega^{-1/2}(\bar{E})f(\bar{E},\omega)r_{mn}\  ,
\end{equation}
where $\omega = E_m-E_n$, $\bar{E} = (E_m + E_n)/2$. 
$\Omega(\bar{E})$ is the density of states, $O(\bar{E})$ and 
$f(\bar{E},\omega)$ 
are smooth functions, and the $r_{mn} = r_{nm}^\ast$ are assumed to be pseudo random numbers that follow a Gaussian distribution with zero mean and unit variance.

From now on, we focus on operators with their smooth diagonal part subtracted, denoted by ${\cal O}$, where 
\new{\begin{equation}\label{def-O}
{\cal O}_{mn} = \langle m |{\cal O} |n\rangle = {O}_{mn} - O(\overline{E})\delta_{mn}  .
\end{equation}
In quantum many-body systems with local interactions, if the observable $\cal O$ satisfies the ETH, one expects that $f(\overline{E},\omega) \approx \text{const.}$ holds within 
certain energy scale (more detailed discussion of such energy scale is presented at the end of this section.).
% energy scales $\Delta E \le \Delta E_{\text{eq}}$.
Therefore, one has ${\cal O}_{mn} \propto r_{mn}\Omega^{-1/2}(\bar{E})$. 
If $r_{mn}$ are truly independent Gaussian random numbers, and $\Omega(\bar{E})$ is approximately constant in this region, then the truncated operator ${\cal O}_{\Delta E}$ would correspond to a sample of random matrices drawn from the Gaussian unitary ensemble (GUE).}

However, since ${\cal O}_{mn}$
is fully determined by the Hamiltonian and the observable, its matrix elements are not independent but instead contain intrinsic correlations \cite{Dymarsky22-ETH-offdiag,Jonas-offdiag,GETH-Foini19}.
Such correlations are crucial in the study of 
multiple time (beyond two) correlation functions, e.g., out-of-time-order correlators \cite{goold-otoc}. 
Furthermore, It has been shown that these correlations persist down to energy scales much smaller than $\Delta E_{\text{eq}}$~\cite{Wang22-ETH-offdiag}. A more recent work further implies that, in a strict sense, the GUE random-matrix description cannot be applied at any finite energy scale $\Delta E \neq 0$ \cite{iniguez2023microcanonical}.%~\cite{iniguez2023microcanonicalmicrocanonical-offdiag}.

The inapplicability of a strict GUE description does not exclude a more general RMT description that allows for correlations between matrix elements. 
Inspired by the 
full ETH \cite{GETH-Foini19,PhysRevLett.129.170603-feth-fp}, a new picture was suggested. It conjectures \cite{wang2024emergence-uie-offdiag} that 
below a sufficiently small energy scale $\Delta E_U$,  ${\cal O}_{\Delta E}$ can be regarded as a sample of unitarily invariant Ensemble (UIE), 
\begin{equation}\label{eq-uie}
    {\cal O}_{\Delta E_U}\equiv{\cal O}_{U}=U{\cal O}^{*}U^{\dagger},
\end{equation}
where $U$ is a Haar-random unitary (or orthogonal) operator and ${\cal O}^\ast$ is some fixed, operator-specific matrix which can be chosen to be diagonal without loss of generality.
From a more conceptual point of view, the above conjecture is built on the picture that eigenstates in a sufficiently narrow energy window are statistically indistinguishable, and thus reshuffling them leaves statistical properties of ${\cal O}_{mn}$ unchanged.
In this framework, when $\Delta E \le \Delta E_U$, the truncated operator ${\cal O}_{\Delta E}$ can be effectively viewed as an operator with fixed eigenvalues and random eigenvectors.

It is important to note that the UIE represents a more general random-matrix ensemble that allows for correlations between matrix elements, and the GUE may be viewed as a special limiting case of the UIE.

\new{The UIE description set strong constraints on the n-point matrix-element correlation functions captured/in the framework of the full
ETH \cite{PhysRevLett.134.140404-feth-fp}. 
For one- and two-point correlation functions, the constraints are already contained in the conventional ETH ansatz.
Concretely, if ${\cal O}_{\Delta E}$ admits a UIE description within $\Delta E \le \Delta E_U$, 
\begin{subequations}\label{eq-O-f12}
\begin{gather}
\langle{\cal O}_{mm}\rangle_{E_{m}\approx E}=\langle O_{mm}\rangle_{E_{m}\approx E}-O(E)={\rm 0}\ , \label{eq-O-f1} \\
\langle{\cal O}_{mn}{\cal O}_{nm}\rangle_{E_{m}\approx E_{1},E_{n}\approx E_{2}}=\Omega^{-1}(\overline{E})f(\overline{E},\omega)={\rm const.}\ , \label{eq-O-f2}
\end{gather}
\end{subequations}
where $\overline{E} = (E_1 + E_2)/2$ and $\omega = E_2 - E_1$.
Here $\langle \cdot \rangle$ denotes an average over nearby eigenstates within an small energy interval $\delta e \ll \Delta E_U$. 
Note that Eq.~\eqref{def-O} has been used in deriving Eq.~\eqref{eq-O-f1}. }
% We further emphasize that Eqs.~\eqref{eq-O-f1} and \eqref{eq-O-f2} constitute necessary conditions for ${\cal O}_{\Delta E}$ to admit a UIE description.

\new{
If $\Omega(\overline{E})$ can be approximately regarded as an constant within $\Delta E \le \Delta E_U$, 
Eq.~\eqref{eq-O-f2} implies that
\begin{equation}
  f(\overline{E},\omega) = {\rm const},\ \ \forall |\omega| \le \Delta E_U\ .
\end{equation}
Since the envelope function $f(\overline{E},\omega)$ is related to the Fourier transform of the thermal auto-correlation function \cite{dkpr-16},
\begin{equation}
C_{\beta}(t)
=
\frac{\text{Tr}\left[e^{-\beta H/2}{\cal O}(t)e^{-\beta H/2}{\cal O}\right]}
{\text{Tr}\left[e^{-\beta H}\right]},
\end{equation}
the plateau width of $f(\overline{E},\omega)$ is set by the inverse equilibration time of this two-point correlation function. 
We denote this energy scale by $\Delta E_{{\rm eq}}^{(2)}$. 
If the equilibration time of $C_{\beta}(t)$ is denoted by $T_{\rm eq}^{(2)}$, then
\begin{equation}
\Delta E_{{\rm eq}}^{(2)} \sim \frac{2\pi}{T_{\rm eq}^{(2)}} .
\end{equation}
The preceding discussion then implies the hierarchy of energy scales
\begin{equation}
    \Delta E_U \leq \Delta E^{(2)}_{\rm eq}.
\end{equation}
For observables coupled to slowest transport mode, 
$\Delta E^{(2)}_{\rm eq}$ coincides with the Thouless energy $\Delta E_{\rm th}$ \cite{altshuler1986repulsion,PhysRevLett.123.210603,PhysRevX.12.021009,PhysRevB.105.104509,winer2023emergent,winer2023reappearance,Roy_Thouless20,Roy_Thouless22,Roy_Thouless23},
which leads to 
\begin{equation}
    \Delta E_{U}\le\Delta E_{{\rm th}}\sim \Delta E_{{\rm eq}}^{(2)}\ .
\end{equation}
}
\subsection{Criteria for the unitarily invariant ensemble description}
After introducing the conceptual framework, we now turn to the question of how to characterize the UIE. 
In particular, we aim at establishing a checkable criterion for identifying an operator as a sample of such an ensemble.

Before going to the details, it is useful to first introduce the notion of free cumulants, which will be used throughout this paper. They are defined in terms of the moments of an operator ${\cal O}$,
${\cal M}_k \equiv \frac{1}{d}\mathrm{Tr}[{\cal O}^k]$~\cite{PhysRevLett.134.140404-feth-fp}, through the recursive relation
\begin{equation}\label{eq-cumulant-moment}
\Delta_{k}
= {\cal M}_{k}
- \sum_{j=1}^{k-1}
\Delta_{j}
\sum_{a_{1} + a_{2} + \cdots + a_{j} = k-j}
{\cal M}_{a_{1}} \cdots {\cal M}_{a_{j}} .
\end{equation}
\new{For example, in the case of operators with a vanishing first moment, $\text{Tr}[{\cal O}]=0$, which is the case considered here, the first six free cumulants read
\begin{subequations}\label{def-fc}
\begin{align}
\Delta_1 &= {\cal M}_1 = 0, \label{eq-free-cumulants-first-six-a}\\
\Delta_2 &= {\cal M}_2, \label{eq-free-cumulants-first-six-b}\\
\Delta_3 &= {\cal M}_3, \label{eq-free-cumulants-first-six-c}\\
\Delta_4 &= {\cal M}_4 - 2{\cal M}_2^2, \label{eq-free-cumulants-first-six-d}\\
\Delta_5 &= {\cal M}_5 - 5{\cal M}_2{\cal M}_3, \label{eq-free-cumulants-first-six-e}\\
\Delta_6 &= {\cal M}_6 - 6{\cal M}_2{\cal M}_4 - 3{\cal M}_3^2 + 7{\cal M}_2^3 .
\end{align}
\end{subequations}
}

If ${\cal O}_{\Delta E_U}$ admits a UIE description \eqref{eq-uie}, 
then statistical properties of its submatrices ${\cal O}_{\delta E} = P_{\delta E} {\cal O}_{\Delta E_U} P_{\delta E}$
 are unambiguously determined by ${\cal O}^*$ and the ratio $d_{\delta E}/d_{\Delta E_U}$, where $d_{\delta E}=\text{Tr}[P_{\delta E}]$ and $d_U \equiv d_{\Delta E_U}=\text{Tr}[P_{\Delta E_U}]$. 
In terms of free cumulants, a compact relation can be derived \cite{wang2024emergence-uie-offdiag,wang2025eigenstate-uie-classical}, 
\begin{equation}\label{eq-fca}
    \Delta_{k}(\delta E)=\left(\frac{d_{\delta E}}{d_{U}}\right)^{k-1}\Delta_k^U,\  \delta E \le \Delta E_U ,
\end{equation}
where $\Delta_k(\delta E)$ and  $\Delta^U_k$  denote the $k$th free cumulant of  ${\cal O}_{\delta E}$ and ${\cal O}_{\Delta E_U}$, respectively.
% For $k=1$ and $k=2$, the scaling relation in Eq.~\eqref{eq-fca} follows directly from Eq.~\eqref{eq-O-f12}, together with the definition in Eq.~\eqref{def-fc}:
% \begin{gather}
% \Delta_{1}(\delta E)=\frac{1}{d_{\delta E}}\sum_{|E_{m}-E_{0}|\le\delta E/2}{\cal O}_{mm}\sim{\rm const},
% \nonumber \\
% \Delta_{2}(\delta E)=\frac{1}{d_{\delta E}}\sum_{|E_{m,n}-E_{0}|\le\delta E/2}|{\cal O}_{mn}|^{2}\sim d_{\delta E}\ .
% \end{gather}
\new{For the sake of a self-contained presentation, we provide a derivation of Eq.~\eqref{eq-fca} in Appendix~\ref{apx-fc}, following the approach of Ref.~\cite{wang2025eigenstate-uie-classical}. }

If $\Delta E_U$ is sufficiently small such that the density of states within the microcanonical window is approximately constant, one has $d_{\delta E}/d_U = \delta E / \Delta E_U$.
In this case, Eq.~\eqref{eq-fca} becomes
\begin{equation}\label{eq-Deltak-DE}
\Delta_{k}(\delta E)
= \left(\frac{\delta E}{\Delta E_U}\right)^{k-1} \Delta_k^{U}
\propto \delta E^{k-1}.
\end{equation}
This scaling relation provides the main criteria for determining whether microcanonically truncated operators admit a UIE description.
\new{To make the scaling relation in Eq.~\eqref{eq-Deltak-DE} more transparent, let us consider the case $k=2$. 
In this case, Eq.~\eqref{eq-Deltak-DE} reduces to
\begin{equation}
    \Delta_{2}(\delta E) \propto \delta E .
\end{equation}
This linear dependence  is expected when the energy window lies within the plateau regime of $f(\overline{E},\omega)$, which can be seen as follows:
\begin{align}
    \Delta_{2}(\delta E) 
    &= \frac{1}{d_{\delta E}}\sum_{|E_{m,n}-E_{0}|\le\delta E/2}|{\cal O}_{mn}|^{2} \nonumber \\
    &\simeq 
    \frac{1}{d_{\delta E}}
    \sum_{|E_{m,n}-E_{0}|\le\delta E/2}
    \Omega^{-1}(\overline{E}) f(\overline{E},\omega) . \label{eq:fw3}
\end{align}}
\jz{If $f(\bar E,\omega)$ and $\Omega(E)$ are approximately
constant within the energy window $|E_{m,n}-E_0|\leq \delta E/2$,
Eq.~\eqref{eq:fw3} gives
\begin{equation}
\Delta_2(\delta E)
\sim
\frac{(\delta E)^2\Omega(E_0)}{d_{\delta E}}
\sim
\delta E ,
\tag{22}
\end{equation}
where $d_{\delta E}\simeq \Omega(E_0)\delta E$ has been used.}

%If $f(\overline{E},\omega)\approx {\rm const.}$ and $\Omega(\overline{E}) \approx {\rm const.}$  in the energy window $|E_{m,n} - E_0| \le \delta E/2$, then
%\begin{equation}
%    \Delta_{2}(\delta E)\sim d_{\delta E}\sim\delta E\ .
%\end{equation}}

Before closing this section, it should be noted that, similar to moments, free cumulants can also be evaluated efficiently using quantum-typicality–based numerical methods \cite{Robin_DQPT_2014, Robin_diffusion_2015, Robin_diffusion_2017,PhysRevLett.102.110403-DQT,Jin2021,HeitmannRichterSchubertSteinigeweg-DQT}, enabling us to access system sizes well beyond the reach of exact diagonalization (ED), as will be explained in more detail in the next section.

%------------------------------------------------------------------------------
% Numerical Investigation
%------------------------------------------------------------------------------
\section{Numerical Investigation }\label{numerical-section}
In this section, we investigate the existence of the UIE description of microcanonically truncated operators for specific  
models through numerical simulations.

%------------------------------------------------------------------------------
% Model and Observable
%------------------------------------------------------------------------------
\subsection{Model and observable}\label{model}

The \jz{first} model we consider here is the Heisenberg XXZ chain described by the Hamiltonian
\begin{equation}\label{eq: integrable system}
H=J\sum_{r=1}^{L}(S_{r}^{x}S_{r+1}^{x}+S_{r}^{y}S_{r+1}^{y}+\Delta S_{r}^{z}S_{r+1}^{z}+\Delta^{\prime}S_{r}^{z}S_{r+2}^{z}),
\end{equation}
where $S_r^\mu({\mu=x,y,z})$ are the components of a spin operator with spin quantum number $S$ at lattice site $r$, and we set $J = 1$. Here $L$ is the system size, and we employ periodical boundary condition (PBC) $S_{r+L}^{\mu}=S_{r}^{\mu}$. \new{For the observable, we focus on the total spin current operator 
\begin{equation}\label{eq: spin current heisenberg}
J_{S}=\sum_{r=1}^{L}j_{r}=\sum_{r=1}^{L}(S_{r}^{x}S_{r+1}^{y}-S_{r}^{y}S_{r+1}^{x}),
\end{equation}
which plays a central role in the study of spin transport  \cite{Bertini-21-RMP-transport}.}

% In the following, we focus on spin quantum numbers $s=\frac{1}{2}, 1, \frac{3}{2}$.

The system is non-integrable for spin quantum number $S > \tfrac{1}{2}$.
For $S=\tfrac{1}{2}$ and $\Delta=\Delta'=0$, the model is integrable and
can be mapped onto a free-fermion model.
When $\Delta \neq 0$ and $\Delta'=0$, the system is integrable in terms of the Bethe ansatz, often referred to as an interacting integrable model.

Note that the system possesses several symmetries.
First of all, the magnetization along the $z$ direction $S^z = \sum_rS_r^z$ is conserved, i.e.~$[H,S^z]=0$. Second, due to PBC, the system is translationally invariant, i.e.~ $[H, {\cal T}] = 0$, where $T$ is the translation operator
\begin{equation}
    {\cal T}S_{r}^{\mu}{\cal T}^{-1}=S_{r+1}^{\mu},\ \ \ \mu=x,y,z.
\end{equation}
One can introduce the quasi-momentum operator $\cal K$ by writing $\cal T$ as
\begin{equation}
    {\cal T} = e^{i{\cal K}}, 
\end{equation}
and the corresponding eigenvalues of $\cal K$ are
\begin{equation}
    k = \frac{2\pi n}{L},\ n = 0,1,\ldots, L - 1.
\end{equation}
The total spin current operator $J_S$ also commutes with $S^z$ and $\cal T$, i.e.~$[J_{S},S^z]=0$ and  $[J_{S},{\cal T}]=0$.

In the remainder of this work, we focus on the subsector with zero magnetization ($S^z = 0$) and zero quasi-momentum ($k = 0$). For large $L$, the Hilbert space dimension of the subsector scales as
\begin{equation}
    d\sim L^{-3/2}D,
\end{equation}
where $D = (2s + 1)^L$ is the dimension of the whole Hilbert space. By exploiting the system's symmetries, the Hilbert-space dimension is substantially reduced, allowing us to access larger system sizes.

Note that, in this subsector, the system possesses additional symmetries, such as spatial reflection symmetry (with respect to the center of the chain) and ``particle–hole'' symmetry. However, the total spin current operator is anti-symmetric under these symmetry transformations. These additional symmetries are not further exploited.

\begin{figure}[tb]
	\centering
\includegraphics[width=1.0\linewidth]{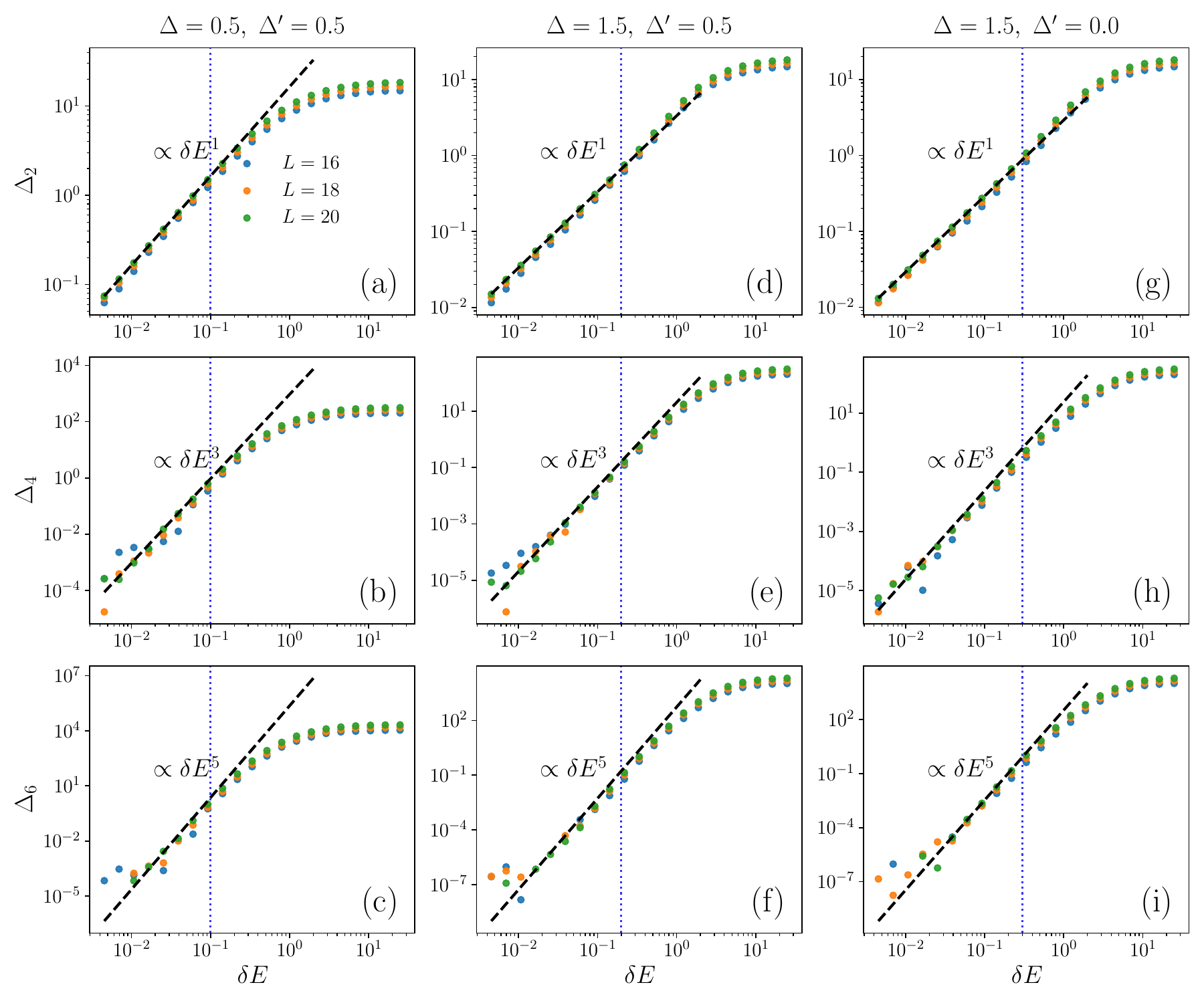}
 \caption{Numerical results for the XXZ model for spin quantum number $S = 1$. Even cumulants $\Delta_k$ for $k=2,4,6$ as a function of $\delta E$ for the total spin current operator $J_S$ for parameter sets [(a)(b)(c)] $\Delta = 0.5, \Delta^\prime = 0.5$;  [(d)(e)(f)] $\Delta = 1.5, \Delta^\prime = 0.5$ and  [(g)(h)(i)] $\Delta = 1.5, \Delta^\prime = 0.0$, for system sizes $L=16, 18, 20$. 
 As a guide to the
eye, the dashed lines and vertical dotted lines 
indicate the scaling $\Delta_{k}\propto\delta E^{k-1}$ and an approximate location of $\Delta E_U$, respectively.
 }\label{Fig-Spin1}
\end{figure}

%------------------------------------------------------------------------------
% Numerical Method
%------------------------------------------------------------------------------
\subsection{Numerical methods}\label{method}
According to Eq.~\eqref{eq-cumulant-moment}, the free cumulants $\Delta_k$ are determined unambiguously by the moments ${\cal M}_{n \le k}$. Therefore, in practice we calculate the moments first. In general, the calculation of moments of the microcanonically truncated operator
\begin{equation}
{\cal M}_{k}(\delta E)=\frac{1}{d_{\delta E}}\text{Tr}[(P_{\delta E}OP_{\delta E})^{k}]
\end{equation}
would require full exact diagonalization of the Hamiltonian. 
In order to access larger system sizes, we employ an approach based on quantum typicality \cite{wang2024emergence-uie-offdiag}, consisting of two main ingredients that will be detailed in the following.

The first ingredient is to approximate the trace of an operator by its expectation value with respect to a normalized Haar-random state $|\psi_{\text{typ}}\rangle$, i.e.,
\begin{equation}\label{eq-M-typ}
{\cal M}_{k}(\delta E)=\langle\psi_{\text{typ}}|(P_{\delta E}OP_{\delta E})^{k}|\psi_{\text{typ}}\rangle+\varepsilon_{\text{typ}},
\end{equation}
with an error $\varepsilon_{\text{typ}}\sim d_{\delta E}^{-1/2}$. The error can be further reduced by averaging over different realizations of Haar-random states, scaling as $\varepsilon_{\text{typ}}\sim(N d_{\delta E})^{-1/2}$, where $N$ denotes the number of the Haar-random states.

The second ingredient is to expand the projection operator in terms of Chebyshev polynomials
\begin{equation}\label{eq-PE}
    P_{\delta E}(H)=\sum_{k=0}^{\infty}C_{k}T_{k}(\frac{H-b}{a}),
\end{equation}
where $a=(E_{\text{max}}-E_{\text{min}})/2$, $b=(E_{\text{max}}+E_{\text{min}})/2$, with $E_{\text{max}}$/$E_{\text{min}}$ being the lower/upper bound of the spectrum.

The coefficients $C_k$ are written as
\begin{gather}
C_{0}=\frac{1}{\pi}\left[\arcsin\left(\frac{\frac{\delta E}{2}+b-E_{0}}{a}\right)-\arcsin\left(\frac{-\frac{\delta E}{2}+b-E_{0}}{a}\right)\right] 
\end{gather}
and  
\begin{align}\label{eq-Ck}
	C_k(k\ge 1)  & = \frac{2(-1)^{k+1}}{\pi k}\biggl[ 
	\sin\biggl(k\arccos\biggl(\frac{\frac{\delta E}{2}+b-E_0}{a}\biggl)\biggl) \nonumber 
	\\
	& -\sin\biggl(k\arccos\biggl(\frac{-\frac{\delta E}{2}+b-E_0}{a}\biggl)\biggl)\biggl] 
	\ .
\end{align}
Practically, we truncate the infinite series in Eq.~\eqref{eq-PE} to a finite order $N_\text{tr}$,
\begin{equation}
    P_{\delta E}(H)\simeq P_{\delta E}^{\text{tr}}(H)\equiv\sum_{k=0}^{N_{\text{tr}}}C_{k}T_{k}(\frac{H-b}{a}),
\end{equation}
such that
\begin{equation}
    {\cal M}_{k}(\delta E)=\langle\psi_{\text{typ}}|\left(P_{\delta E}^{\text{tr}}(H){\cal O}P_{\delta E}^{\text{tr}}(H)\right)^{k}|\psi_{\text{typ}}\rangle+\varepsilon_{\text{typ}}+\varepsilon_{\text{tr}}. 
\end{equation}
The truncation error $\varepsilon_{\text{tr}}$ decreases with increasing $N_{\text{tr}}$, while the numerical cost grows linearly with $N_{\text{tr}}$ at the same time. In our numerical simulations, we choose
$N_{\text{tr}} = 10a\,\frac{2\pi}{\Delta E}$.

\begin{figure}[tb]
	\centering
\includegraphics[width=1.0\linewidth]{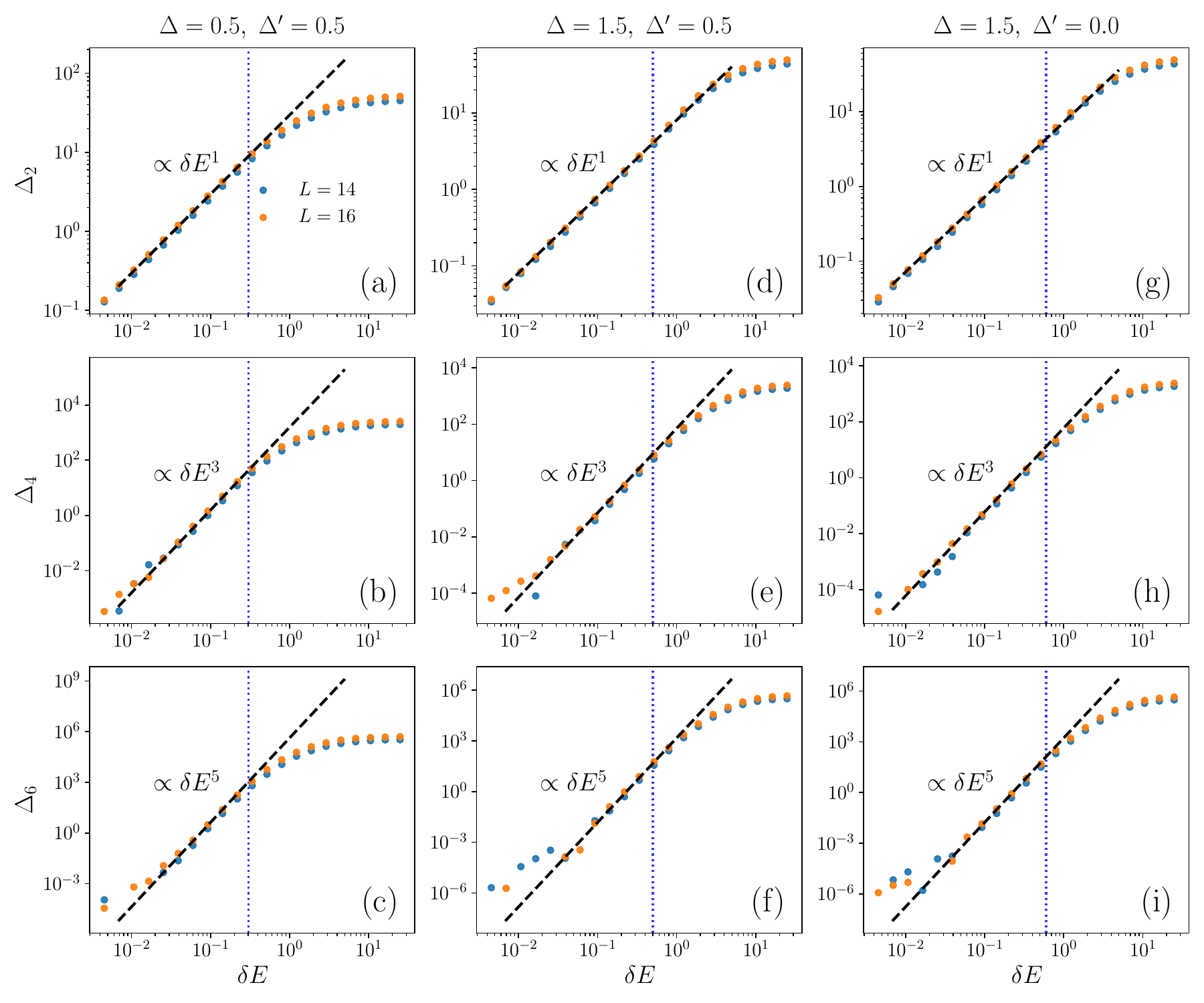}
 \caption{Similar data as the one in Fig.~\ref{Fig-Spin1} but for spin quantum number $S = 3/2$.
 }\label{Fig-Spin32}
\end{figure}

The computationally most expensive part of the simulation is the evaluation of
\begin{equation}
P_{\delta E}^{\text{tr}}(H)|\psi\rangle=\sum_{k=0}^{N_{\text{tr}}}C_{k}T_{k}(\frac{H-b}{a})|\psi\rangle .
\end{equation}
It can be carried out efficiently by making use of the recurrence relation of Chebyshev polynomials and sparse matrix--vector multiplications \cite{Jin2021}.
With this approach, we are able to access
system sizes up to $L = 30, 20,$ and $16$ for spin quantum numbers
$S = 1/2, 1,$ and $3/2$, respectively.

%------------------------------------------------------------------------------
% Results
%------------------------------------------------------------------------------
\subsection{ Results}\label{numerics}
After having introduced the model, observables, and numerical methods, this section presents numerical results obtained for the spin-$S$ Heisenberg XXZ model.
Our main goal is to demonstrate the existence of a UIE description for microcanonically truncated operators for different spin quantum numbers $S=\frac{1}{2},1,\frac{3}{2}$  and parameter regimes. We do so by numerically verifying Eq.~\eqref{eq-Deltak-DE}. 

In Fig~\ref{Fig-Spin1}, we present results for $S = 1$, where the free cumulants $\Delta_k$ are shown as a function of $\delta E$. 
Since the total spin current operator $J_S$ is anti-symmetric under spatial reflection, all odd moments and cumulants vanish, and therefore only even cumulants are shown here.
We consider three different parameter sets, 
$\Delta = 0.5, \Delta^\prime = 0.5$, $\Delta = 1.5, \Delta^\prime = 0.5$ and $\Delta = 1.5, \Delta^\prime = 0.0$ and similar results are seen across all these cases. Below an energy scale $\Delta E_U$, the power-law scaling $\Delta_k\propto \delta E^{k-1}$ can be observed, consistent with Eq.~\eqref{eq-Deltak-DE}.
Deviations from the power-law scaling are found when $ \delta E$ is extremely small. However, these deviations tend to shift to smaller values of $\delta E$ for larger system sizes, suggesting that the scaling $\Delta_k \propto \delta E^{k-1}$ may persist down to arbitrarily small energy windows in the thermodynamic limit.
The results provide evidence for supporting the conjecture that within a sufficiently small energy scale $\Delta E \le \Delta E_U$, the microcanonically truncated operator ${\cal O}_{\Delta E}$ admit an UIE description.

\begin{figure}[tb]
	\centering
\includegraphics[width=1.0\linewidth]{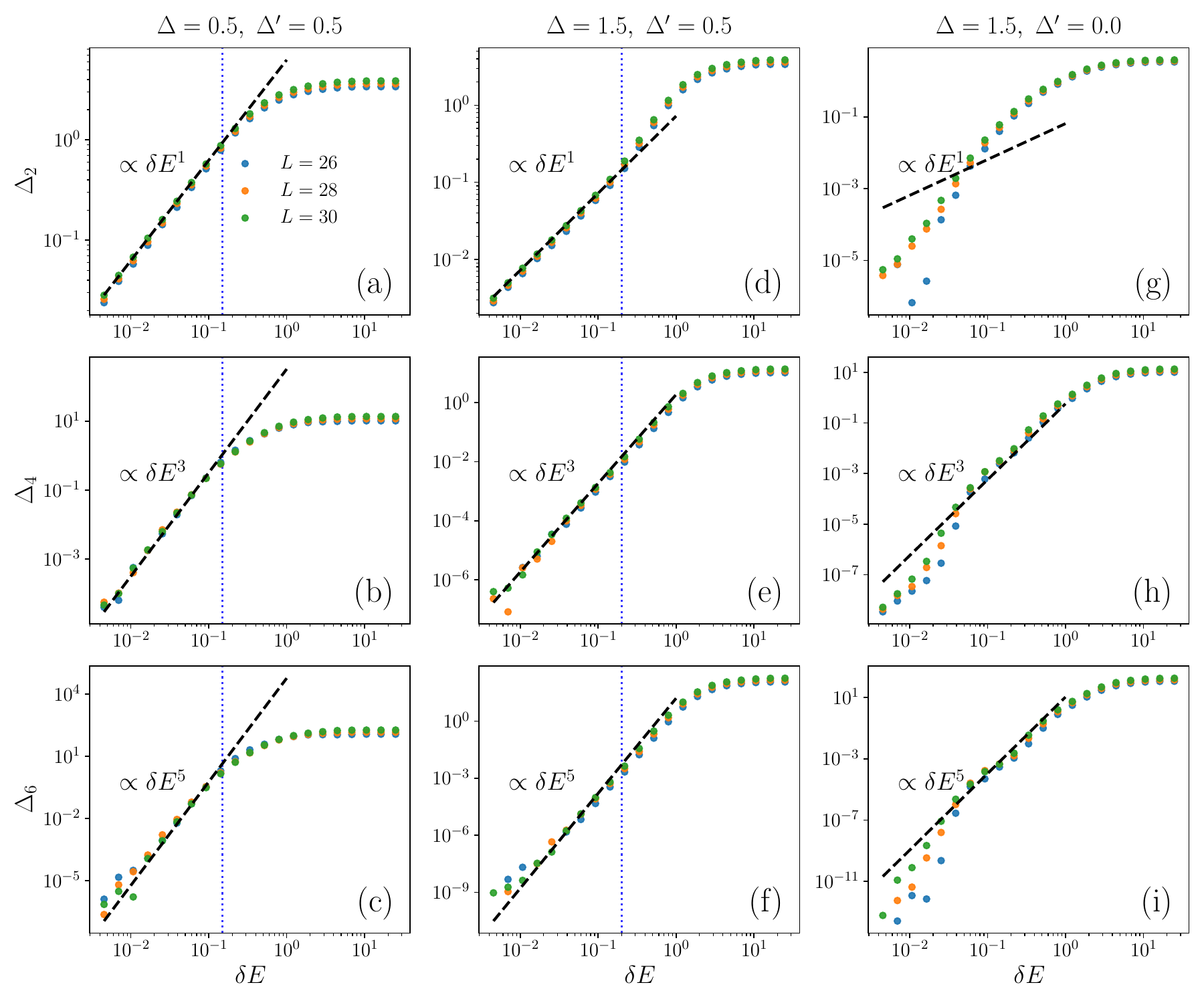}
 \caption{Similar data s the one in Fig.~\ref{Fig-Spin1} but for spin quantum number $S = 1/2$.
 }\label{Fig-Spin12}
\end{figure}

Similar results are observed for  $S = 3/2$ and for the chaotic cases of $S = 1/2$ ($\Delta = 0.5,\ \Delta^\prime = 0.5$ and $\Delta = 1.5,\ \Delta^\prime = 0.5$), as shown in Fig.~\ref{Fig-Spin32} and Fig.~\ref{Fig-Spin12}[(a)–(f)] respectively. In these cases, the power-law scaling $\Delta_k \propto \delta E^{k-1}$ is present within a certain energy scale.
In contrast, for $\Delta = 1.5,\ \Delta^\prime = 0.0$, where the system is integrable, no clear power-law scaling is observed within the energy scale considered here, suggesting the absence of a UIE description for the microcanonically truncated operator.
It should be noted that the deviation of the second cumulant from linear scaling, $\Delta_{2}(\delta E) \not\propto \delta E$, indicates the absence of a plateau in the envelope function $f(\overline{E}, \omega)$, despite the fact that the system exhibits diffusive magnetization transport. This discrepancy can be attributed to the known anomalous scaling of conductivity
found in integrable fermion systems \cite{PhysRevB.70.205129-current-int,PhysRevE.87.012118-current-int,PhysRevLett.107.250602-current-int}.

In summary, the numerical results presented above provide evidence in support of the conjecture that microcanonically truncated operators exhibit RMT universality in chaotic systems, whereas no clear signature of such behavior is observed in integrable cases.

\section{Conclusion and Discussion}\label{conclusion}
In this paper, we numerically investigated the conjecture that the statistical properties of microcanonically truncated operators exhibit RMT universality within sufficiently narrow energy windows. In particular, 
we focused on the total spin current operator in translationally invariant spin-$S$ Heisenberg chains with $S=\frac{1}{2}, 1, \frac{3}{2}$, using a quantum-typicality-based numerical approach combined with exploitation of symmetries.

Our results provide evidence that, in chaotic regimes and for sufficiently small energy windows, microcanonically truncated operators admit an UIE description. In contrast, no clear signature of such a description is observed in integrable cases. These findings support the conjecture that RMT universality in observable statistics is a general feature of quantum-chaotic many-body systems.

%------------------------------------------------------------------------------
% Acknowledgement
%------------------------------------------------------------------------------
\section{Acknowledgements}
This work has been funded by the Deutsche Forschungsgemeinschaft (DFG), under Grant No. 531128043, as well as under Grant No. 397107022, No. 397067869, and No. 397082825 within the DFG Research Unit FOR 2692, under Grant No. 355031190. We additionally acknowledge computing time on HPC3 cluster under DFG Grant No. 456666331.

\section{Data availability} 
The data that support the findings of this article are openly available \cite{kempa_2026_20343383}.

\appendix
\section{Derivation of Eq.~\eqref{eq-fca}}\label{apx-fc}

\new{In this appendix, we give the derivation of Eq.~\eqref{eq-fca}, following the approach of Ref.~\cite{wang2025eigenstate-uie-classical}.
We begin by recalling the UIE introduced in Eq.~\eqref{eq-uie},
\begin{equation}
{\cal O}_{U}=U{\cal O}^{*}U^{\dagger},
\end{equation}
where $U$ is a Haar-random unitary operator. 
We take ${\cal O}$ as a typical sample of ${\cal O}_U$, in a Hilbert space of dimension $d_U$. In a given basis, the $k$th cumulant can be written as
\begin{equation}\label{eq-Deltak-Oab}
\Delta_{k}
=
\frac{1}{d_U}
\sum_{m_{1}\neq m_{2}\neq\cdots\neq m_{k}}
{\cal O}_{m_{1}m_{2}}
{\cal O}_{m_{2}m_{3}}
\cdots
{\cal O}_{m_{k}m_{1}},
\end{equation}
where ${\cal O}_{mn}=\langle m|{\cal O}|n\rangle$. 
For sufficiently large $d_U$, one may replace the product 
${\cal O}_{m_{1}m_{2}}{\cal O}_{m_{2}m_{3}}\cdots{\cal O}_{m_{k}m_{1}}$
in Eq.~\eqref{eq-Deltak-Oab} by its UIE average. 
According to Ref.~\cite{Maillard_2019_geth},
\begin{equation}\label{eq-O-uim}
\langle{
{\cal O}_{m_{1}m_{2}}
{\cal O}_{m_{2}m_{3}}
\cdots
{\cal O}_{m_{k}m_{1}}
}\rangle_{\rm UIE}
=
\frac{1}{d_U^{k-1}}\Delta_{k}
=
\text{const.}\ .
\end{equation}}

\new{We now truncate ${\cal O}$ to a $d$-dimensional subspace $S_d$ of the Hilbert space by introducing the projection operator
\begin{equation}
P_{d}=\sum_{m\in S_{d}}|m\rangle\langle m| ,
\end{equation}
which gives
\begin{equation}\label{eq-Od}
{\cal O}_{d}=P_{d}{\cal O}P_{d}.
\end{equation}
Similar to Eq.~\eqref{eq-Deltak-Oab}, the $k$th cumulant of ${\cal O}_{d}$ reads
\begin{equation}\label{def-PBRM}
\Delta_{k}(d)
=
\frac{1}{d}
\sum_{\substack{
m_{1}\neq m_{2}\neq\cdots\neq m_{k}\\
m_{1},\ldots,m_{k}\in S_{d}
}}
{\cal O}_{m_{1}m_{2}}
{\cal O}_{m_{2}m_{3}}
\cdots
{\cal O}_{m_{k}m_{1}} .
\end{equation}
Using the same argument and inserting Eq.~\eqref{eq-O-uim}, one obtains
\begin{equation}\label{eq-Od-uim}
\Delta_{k}(d)
=
\frac{1}{d}
\sum_{\substack{
m_{1}\neq m_{2}\neq\cdots\neq m_{k}\\
m_{1},\ldots,m_{k}\in S_{d}
}}
\frac{\Delta_{k}}{d_U^{k-1}}
=
\left(\frac{d}{d_U}\right)^{k-1}\Delta_{k},
\end{equation}
which coincides with Eq.~\eqref{eq-fca}.}

\section{Integrable spin-1 model}\label{apx-zamo}

In this appendix, we complement the results of Sec.~\ref{numerics} by
investigating an integrable spin-1 model, namely the
Zamolodchikov-Fateev (ZF) model \cite{zamolodchikov1980model}.
We compare our findings with those obtained for the non-integrable spin-1
models discussed in the main text. 
This comparison further highlights the role of non-integrability for the emergence of a UIE description.
\begin{figure}[tb]
    \centering
    \includegraphics[width=0.5\linewidth]{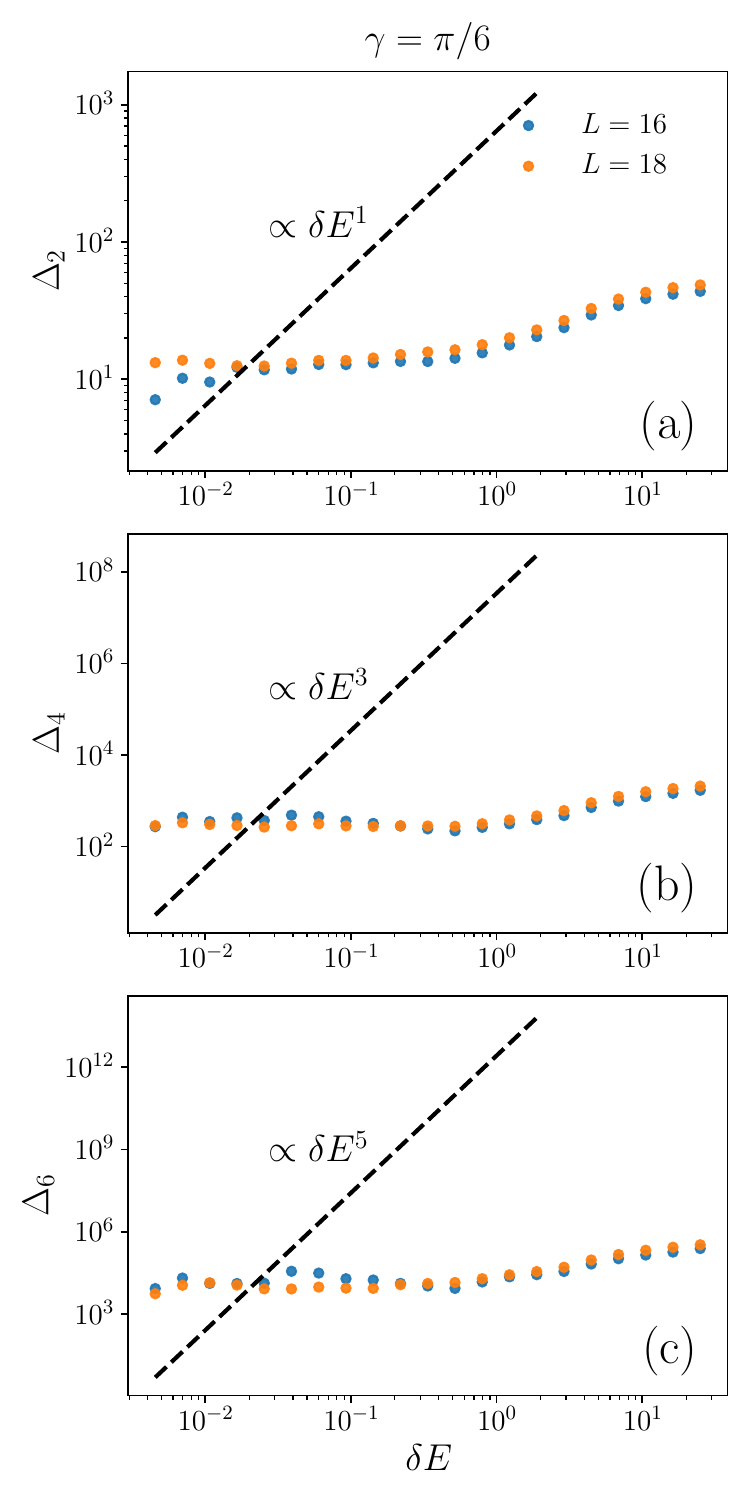}
    \caption{Even free cumulants $\Delta_k$ for $k = 2, 4, 6$ as a function
    of $\delta E$ for the spin current operator $J_\mathrm{ZF}$ of the integrable
    Zamolodchikov-Fateev model with $\gamma = \pi/6$ in Eq.~(\ref{eq: hj-ZF}), for system sizes
    $L = 16, 18$ and $E_0=\mathrm{tr}(H_{\mathrm{ZF}})/d$.
    As a guide to the eye, the dashed lines indicate the scaling
    $\Delta_k \propto \delta E^{k-1}$.
    In contrast to the non-integrable spin-1 results shown in
    Fig.~\ref{Fig-Spin1}, no clear power-law scaling is observed within
    the energy windows considered here.
    }\label{Fig-ZamoS1}
\end{figure}
The Hamiltonian of the ZF model~\cite{Piroli_2016} follows the decomposition $H_{\rm ZF} = \sum_{r=1}^{L} h_r$,
where the local Hamiltonians $h_r$ are given by
\begin{equation}\label{eq: hj-ZF}
\begin{split}
    h_r =\;& S^{x}_{r} S^{x}_{r+1}
           + S^{y}_{r} S^{y}_{r+1}
           + \cos(2\gamma)\, S^{z}_{r} S^{z}_{r+1} \\
          &+ 2\bigl[(S^{x}_{r})^2
           + (S^{y}_{r})^2
           + \cos(2\gamma)\,(S^{z}_{r})^2\bigr] \\
          &- \sum_{a,b} A_{ab}(\gamma)\,
             S^{a}_{r} S^{b}_{r} S^{a}_{r+1} S^{b}_{r+1}\, ,
\end{split}
\end{equation}
with the coefficients $A_{ab}(\gamma) = A_{ba}(\gamma)$ defined by
\begin{align}\label{eq-Aab-ZF}
    A_{xx} &= A_{yy} = 1\,,\nonumber\\
    A_{zz} &= \cos(2\gamma)\,,\nonumber\\
    A_{xy} &= 1\,,\nonumber\\
    A_{xz} &= A_{yz} = 2\cos(\gamma) - 1\,.
\end{align}
Note that, $H_\mathrm{ZF}$ is integrable for any choice of ${\gamma \in \mathbb{R}}$~\cite{Piroli_2016} 
and reduces to the so-called Babujian-Takhtajan model~\cite{Bubujian1982, Takhtajan1982} for the case $\gamma = 0$.

Analogous to Eq.~(\ref{eq: spin current heisenberg}), the spin current operator for the ZF model is obtained via the lattice continuity equation for the local magnetization $S^z_r$, 
i.e. ${J_{\mathrm{ZF}} = \mathrm{i}\sum_{r=1}^{L}\bigl[h_{r-1},\, S^z_r\bigr]}$, 
which yields~\cite{Piroli_2016}
\begin{equation}\label{eq: J-ZF}
    J_{\mathrm{ZF}} = \frac{\mathrm{i}}{2}\sum_{r=1}^{L}
          \Bigl(J^{(1)}_{r-1,r} + J^{(2)}_{r-1,r} + J^{(3)}_{r-1,r}\Bigr)\,,
\end{equation}
where
\begin{align}
    J^{(1)}_{r-1,r} &=
        S^+_{r-1}S^-_r - S^-_{r-1}S^+_r\,,\nonumber\\
    J^{(2)}_{r-1,r} &=
        \bigl(S^-_{r-1}\bigr)^{\!2}\bigl(S^+_r\bigr)^{\!2}
        - \bigl(S^+_{r-1}\bigr)^{\!2}\bigl(S^-_r\bigr)^{\!2}\,,\nonumber \\
    J^{(3)}_{r-1,r} &=
        (1 - 2\cos(\gamma))
        \bigg[\bigl(S^+_{r-1}S^-_r - S^-_{r-1}S^+_r\bigr)S^z_{r-1}S^z_r\, \nonumber\\
       &+ S^z_{r-1}S^z_r\bigl(S^+_{r-1}S^-_r - S^-_{r-1}S^+_r\bigr)\bigg]\,.
        \label{eq-J3-ZF}
\end{align}
Here, $S_r^\pm = S_r^x \pm \mathrm{i}S_r^y$ denote the spin raising and
lowering operators acting on site $r$.

For our numerical investigation, we set $\gamma = \pi/6$.
Note that, qualitatively similar results are obtained for other values of $\gamma$.
As in the main text, all calculations are performed in the subsector with zero total
magnetization ($S^z = 0$) and zero quasi-momentum ($k = 0$). 
%Remark that, due to the spectrum of the $H_{\mathrm{ZF}}$ we choose (in the symmerty subsector) $E_0 =\mathrm{tr}(H_\mathrm{ZF})/d $. -> better in caption

The numerical results for the even free cumulants $\Delta_k$ ($k = 2, 4, 6$) are shown in Fig.~\ref{Fig-ZamoS1}.
In contrast to the non-integrable spin-1 cases discussed in Sec.~\ref{numerics}, no clear power-law scaling
$\Delta_k \propto \delta E^{k-1}$ is observed within the energy windows considered. 
This suggests the absence of a UIE description for the microcanonically truncated spin current 
operator in the integrable ZF model, consistent with the results in the main text. 

\jz{Together with the results for the integrable XXZ chain with
$\Delta=1.5$, $\Delta^\prime=0$ and spin quantum number $S = \frac{1}{2}$ shown in
Fig.~\ref{Fig-Spin12}[(g)--(i)], this indicates that neither integrable
model exhibits the UIE scaling. At the same time, the free cumulants display qualitatively different behaviors in the two cases, which may be related to their distinct transport properties. The XXZ chain in this regime exhibits diffusive magnetization transport
\cite{PhysRevB.70.205129-current-int,PhysRevE.87.012118-current-int,
PhysRevLett.107.250602-current-int}, whereas the ZF model exhibits
ballistic magnetization transport \cite{Piroli_2016}. A more detailed investigation of the connection between the scaling of free cumulants and transport properties represents an interesting direction and is left for future work.}

%Even at the smallest energy window investigated, which covers only approximately $590$ states,
%corresponding to roughly $0.02\%$ of the states in the relevant symmetry sector, 
%no signatures of the expected scaling behavior are visible.
%Thus, probing even smaller energy windows is numerically not meaningful.
%
%In summary, these findings are consistent with the integrable spin-$\frac{1}{2}$
%results discussed in Sec.~\ref{numerics}, and together provide further
%evidence that the UIE description of microcanonically truncated operators
%is a characteristic feature of chaotic, non-integrable quantum many-body
%systems, and is absent in integrable cases.

%\bibliographystyle{apsrev4-2-titles}
%\bibliography{Ref.bib}

\end{document}